\documentclass[prl,twocolumn,aps,amsmath,superscriptaddress,showpacs]
{revtex4}
\usepackage{graphicx}
\newcommand{\beq}{\begin{equation}}
\newcommand{\eeq}{\end{equation}}
\newcommand{\beqa} {\begin{eqnarray}}
\newcommand{\eeqa} {\end{eqnarray}}
\newcommand{\VEC}[1]{{\bf #1}}

\begin{document}
\title{Fluctuation-Induced Casimir Forces in Granular Fluids.}

\author{C.~Cattuto}
\affiliation{Museo Storico della Fisica e Centro Studi e Ricerche
``Enrico Fermi'', Compendio Viminale, 00184 Roma, Italy}
\affiliation{Frontier Research System, The Institute of Physical
and Chemical Research (RIKEN), Saitama, 351-0198, Japan}

\author{R.~Brito}
\affiliation{Departamento de F\'{\i}sica Aplicada I and GISC, Universidad
Complutense, 28040 Madrid, Spain}

\author{U.~Marini Bettolo Marconi}
\affiliation{Dipartimento di Fisica, Universit\`a di Camerino,
Italy}

\author{F.~Nori}
\affiliation{Frontier Research System, The Institute of Physical
and Chemical Research (RIKEN), Saitama, 351-0198, Japan}
\affiliation{Physics Department and MCTP, The University of
Michigan, Ann Arbor, MI 48109-1040, USA}

\author{R.~Soto}
\affiliation{Departamento de F\'{\i}sica Aplicada I and GISC, Universidad
Complutense, 28040 Madrid, Spain} \affiliation{Departamento de
F\'{\i}sica, FCFM, Universidad de Chile, Casilla 487-3, Santiago,
Chile}

%%%%%%%%%%%%%%%%%%%%%%%%%%%%%

\begin{abstract}
We have numerically investigated the behavior of driven
non-cohesive granular media and found that two fixed large
intruder particles, immersed in a sea of small particles,
experience, in addition to a short range depletion force, a long
range repulsive force. The observed long range interaction is
fluctuation-induced and we propose a mechanism similar to the
Casimir effect that generates it: the hydrodynamic fluctuations
are geometrically confined between the intruders, producing an
unbalanced renormalized pressure. An estimation based on computing
the possible Fourier modes explains the repulsive force and is in
qualitative agreement with the simulations.
\end{abstract}
\pacs{
45.70.-n, %%% Granular systems (see also 05.65.+b Self-organized systems)
05.40.-a, %%% Fluctuation phenomena, random processes, noise, and Brownian
          %%% motion
45.70.Mg  %%% Granular flow: mixing, segregation and stratification
}
\maketitle

%%%%%%%%%%%%%%%%%%%%%%%%%%%%%

Granular materials have been extensively investigated
because of their complex dynamics \cite{general}. Examples of this
include: %Lack of equipartition, 
pattern formation, Faraday waves,
avalanches, convection phenomena, segregation and many more. One of the most active
fields is the behavior of granular mixtures, e.g. Brazil nut
effect~\cite{kudrolli} and its multiple
variations~\cite{Brazil-varia}. In these experiments, the granular
material is composed of a mixture of two types of particles
differing in mass or size. When the system is agitated, particles
of different types may group together (demixing) or stay
mixed~\cite{Cattuto}.
Phase diagrams for the mixing/demixing transition have been
constructed for different material properties or experimental
conditions. However, a fundamental question remains unanswered: Is
the mixing or demixing caused by an effective long range force
between the particles? If so, which is its origin? Let us note that
a long range force is difficult to justify a priori, as grains
interact only through a short ranged hard-core potential.

Recently, three experiments on driven granular mixtures
\cite{Aumaitre,Sanders,Kellay}, performed under very different
experimental conditions, give a hint on how to answer the previous
question. These experiments have shown that thermodynamic
properties (like pressure~\cite{Aumaitre}, density~\cite{Sanders}
and velocity fluctuations~\cite{Kellay}) are different in the
regions between the larger particles,
versus the remaining, external, regions. Consequently, the big
particles modify some physical properties in the confined area
between them. The most likely reason is that the larger particles
limit the allowed wavevectors of the hydrodynamic fluctuations of
the small particles that surround the larger ones.

Forces arising from the confinement of a fluctuation spectrum have
attracted attention since the seminal work of Casimir in 1948, who
predicted the existence of an attractive force between two metal
plates, separated by a vacuum, due to constraints on the quantum
electromagnetic field in the gap imposed by the conducting
plates~\cite{CasimirOriginal}. In fact, the concept of Casimir
force is more general and is common to systems characterized by
long-range fluctuations subject to a geometrical constraint which
limits the long-wavelength portion of their spectrum. Soft
condensed matter provides examples of Casimir forces, such us
those arising in confined critical fluids, in liquid crystals and
superconducting films, where long-range correlations are the
consequence of a broken continuous symmetry~\cite{Kardar}.
Vibrated granular matter~\cite{vNE} and granular
avalanches~\cite{Bretz} provide other examples of physical system
where correlations can become long ranged, in spite of having
short-range forces.

In this Letter we show that there is an effective long range force between
the large and heavy particles in a granular mixture. 
Two ingredients are required: (i) long range correlations and
(ii)  the confinement of the fluctuation spectrum induced
by the large particles in the density, velocity and temperature fields. 

We consider the driven granular model in \cite{PengOtha,vNE}.
Grains are hard particles of diameter $d$ and mass $m$ and their
collisions are characterized by a constant normal restitution
coefficient $\alpha$. To achieve a stationary state, energy is
supplied into the system by random forces acting on all particles.
The random forces $\VEC{F}_i$ are modeled as a white noise of
intensity $\Gamma$: $\langle \VEC{F}_i(t) \VEC{F}_k(t')\rangle =
m\Gamma \delta_{ik}\delta(t-t')$. 
This system reaches a homogeneous stationary state characterized by long range correlations, 
leading to the
renormalization of the energy density and collision frequency due
to the fluctuations at low wavevectors~\cite{vNE}.

The system is composed of $N$ grains put in a square box with
periodic boundary conditions. Besides the small grains, two
inelastic impenetrable and immobile large hard disks (the
intruders) of diameter $D$ are placed, separated at a distance
$R$. The coefficient of restitution $\alpha$ is the same for all
types of collisions.
 This granular mixture is studied using molecular dynamics simulations. The
grain-grain and grain-intruder collisions are treated as usual,
using an event-driven code. To take into account the random
forces, a new type of event is introduced: collisions between
particles and a ``thermal bath''. The event results in a momentum
$\VEC{p}$ being instantaneously transferred to the particle, where
$\VEC{p}$ is randomly chosen by sampling a Gaussian distribution
with zero mean and given variance $P_{\rm bath}^2$ in each
direction. Interactions with the thermal bath are scheduled for
each grain. When an interaction takes place for a particle, its
momentum gets updated and a new future event is scheduled for the
same particle, after a random interval of time $t_{\rm next}$,
according to an exponential distribution $P(t_{\rm next}) \sim
\exp(-t_{\rm next}/\tau_{\rm bath})$. In the limit $\tau_{\rm
bath}\to 0$ and $P_{\rm bath}\to 0$, this injection method
converges to the white noise force with $\Gamma=P_{\rm
bath}^2/m\tau_{\rm bath}$. In practice, the time-scale for
interaction with the thermal bath, $\tau_{\rm bath}$, is taken
smaller than the free-flight-time.

Hereafter, we choose as basic units $d$, $m$, and $\Gamma$. These
units define the time unit as $t_0=(md^2/\Gamma)^{1/3}$ and energy
unit as $e_0=(md^2\Gamma^2)^{1/3}$. We take the diameter of the intruders
$D=8d$ and the coefficient of restitution
$\alpha=0.8$. We simulate systems of size $L=60 \, d$ and $L=80 \, d$, with
the
number density of the granular fluid $n=0.366 \; d^{-2}$. Given
the density, restitution coefficient and noise intensity, the
stationary temperature can be computed using mean field models
giving $T_0=1.84 \, e_0$, and the collision frequency is
$\nu_0=3.03 \, t_0^{-1}$. Hydrodynamic fluctuations determine a
stationary temperature higher than $T_0$ that depends on the
system size~\cite{vNE}. For $L=60 \, d$, $T=2.43 \, e_0$ and for
$L=80 \, d$, $T=2.46 \, e_0$. For every configuration, simulations
were run for about $5\times10^6$ collisions per particle. We
investigated the effective interaction between the intruders at a
distance $R$. For this purpose, we measured  the component of the
total momentum transferred from the gas to intruders 1 and 2 along
the line, parallel to the x-axis, joining their centers, $P_{i x}$
($i=1,2$) as an average over a time interval $\tau=33.4 \, t_0$
(corresponding to about $100$ collisions per particle). This
procedure gives the ``instantaneous'' value of the fluctuating
force as $F_{12}=\langle P_{2 x}- P_{1 x} \rangle/2 \tau$, whose
time-average finally leads to the net effective force, $F$. In the
elastic case, $\alpha=1, \Gamma=0$, $F$  vanishes as expected, whereas
for
$\alpha<1, \Gamma\neq 0$, $F$ is definitively different from zero,
showing an effective force between intruders. The average $y$-component of
the force is compatible with zero to numerical accuracy.
Varying the noise strength, we checked that the force
is proportional to the granular temperature, as can be
deduced by dimensional analysis.

\begin{figure}[tb]
\includegraphics[angle=0,clip=true,width=0.9\columnwidth,
keepaspectratio]{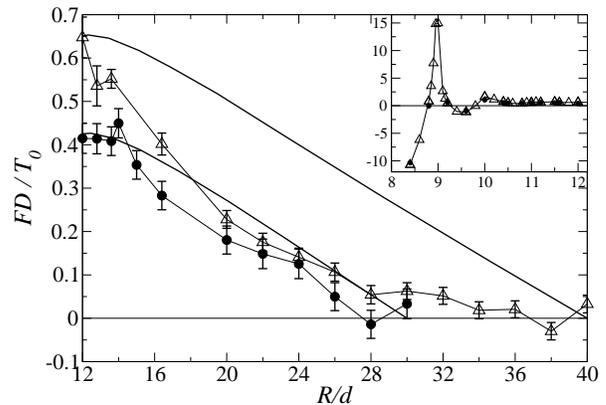} \caption{ Relative effective
dimensionless force $FD/T_0$ between the large particles as
a function of the distance $R$ between their centers. We plot
simulations results for $L=60d$ (open triangles) and for $L=80d$
(filled circles), together with  theoretical predictions (solid
lines). Inset: short range region showing the depletion forces.
Note the difference in the vertical scale, compared with the
long-range part. Error bars in the inset are smaller than the symbols.}
\label{fig.force}
\end{figure}

Figure \ref{fig.force} shows the relative force as a function of
distance $R$ for the two system sizes. The inset shows the short
range part of the relative force, that extends for some small
particle diameters, $d$, alternating between attraction and
repulsion. At larger distances, a repulsive force is observed with
an interaction range much larger than $d$ and comparable with $D$
or the box width (actually, due to the periodic boundary
conditions, the force must vanish at $R=L/2$, as seen from the
simulations).

%%%%%%%%%%%%%%%%% FIGURE 1
%%%%%%%%%%%%%%%%% END OF FIGURE 1 %%%%%%%%%%%%%%%%%%%

The oscillatory short range part is similar to the depletion forces
appearing in fluids that develop local layering
structures~\cite{Evans,DepletionExp}, which are  usually explained
by entropic arguments based on equilibrium statistical mechanics.
Recent works~\cite{Sanders,Kumaran} suggest that 
depletion forces might be at work in granular mixtures, so the concept of
entropic forces is applicable here.
However, the long range part cannot be due to a depletion mechanism
as it extends beyond the typical range of the
depletion forces, and
because it is repulsive in all its range.

\begin{figure}[tb]
 \includegraphics[angle=0,clip=true,width=0.90\columnwidth]
{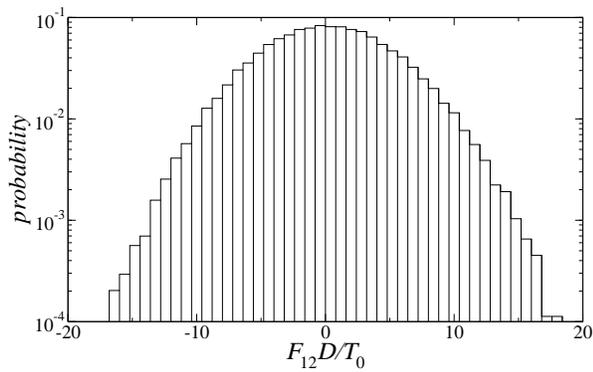}
\caption{Probability distribution of the fluctuating force $F_{12}$ at $R=
20d$, for $L=80d$.
The net effective force is obtained as the average of this histogram. The
average value is $F=0.223 T_0/D$ and the standard deviation is
$\sigma_F=4.83T_0/D$, that is
about $20$ larger than the average.
}
\label{fig.probdistr}
\end{figure}

To elucidate the nature of this force, we analyzed the probability
distribution of the force over the intruders, plotted in 
Fig.~\ref{fig.probdistr}. There it is seen that fluctuations are about $20$
times larger than the average force, and therefore  very long
simulations are required. As the force is proportional
to the granular temperature, these large fluctuations are not an
artifact of a high temperature, but an intrinsic property. Large
fluctuations is a generic signature of fluctuation-induced forces
(see, e.g., ~\cite{Bartolo,Kardar}).

%%%%%%%%%%%%%%%%% FIGURE 2   %%%%%%%%%%%%%%%%%%%%%%%%%%%%%
%%%%%%%%%%%%%%%%% END OF FIGURE 2 %%%%%%%%%%%%%%%%%%%

\begin{figure}[tb]
   \centering
   \includegraphics[width=.75\columnwidth]{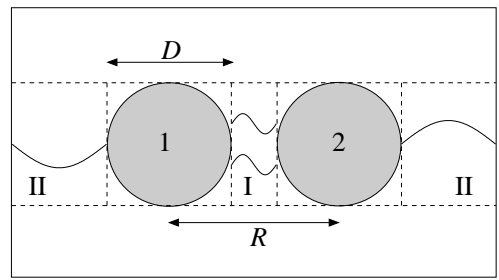}
   \caption{Sketch of the hydrodynamic fluctuations leading to
the Casimir-like force.
In the gap between the intruders only fluctuations of
wavelength
smaller than $R-D$ are allowed, while in the outer space the wavelengths
can be larger.}
   \label{fig.geometry}
\end{figure}

This fact together with the known property of large fluctuations in
driven granular media, suggest that the long range repulsion is a
fluctuation-induced force as in
the Casimir effect~\cite{CasimirOriginal,Kardar}.
The confinement of the hydrodynamic fluctuations between
intruders restricts the allowed fluctuation modes to those with wavelengths
smaller than the gap size, whereas the spectrum of fluctuations in the
outer region allows smaller wavevectors and forms a quasi-continuum,
as illustrated in Fig.~\ref{fig.geometry}.
Consequently, the hydrodynamically-generated ``radiation
pressure'' {\it between\/} the intruders is different from the one
outside this inner region.

To describe the Casimir force originated by the fluctuating
hydrodynamic fields on the intruders we use an approach similar to
Ref.~\cite{vNE}, where the pressure is renormalized in each point
due to fluctuations that are computed using fluctuating
hydrodynamics. Instantaneously, the pressure tensor at position
$\VEC r$ is given by
\beq 
p^*(\VEC r) = p(n(\VEC r),T(\VEC r)) \; I + m \, n(\VEC r) \;
\VEC u(\VEC r) \; \VEC u(\VEC r), 
\eeq
where $n$ and $T$, and $\VEC{u}$ are the fluctuating
density, temperature, and velocity fields and $I$ the identity
tensor, respectively. Moreover, $p(n,T)=T H(n)$ is the usual
thermodynamic pressure for hard disks \cite{verlet-levesque} with
$H(n)=n(1+\phi^2/8)/(1-\phi)^2$, and $\phi=\pi n d^{2}/4$. As the
intruders are immobile, $\VEC{u}$ vanishes at their surface, so
the contribution of the convective term in $p^*$ vanishes and it
becomes a scalar. This would not be the case if the intruders 
were allowed to move. 

Linearizing the hydrodynamic fields $(n,T,\VEC{u})$ around the stationary
values, $(n_0,T_0,0)$, we can expand the pressure
up to second order in the fluctuations $\delta n$ and $\delta T$.
Its statistical average over the random noise is
\beq
\langle p^*\rangle = p_0 + H_1 \langle\delta T \delta n\rangle + T_0H_2
\langle\delta n^2 \rangle ,
\label{p.renorm}
\eeq
where $p_0=p(n_0,T_0)$, $H_1=dH/dn|_{n_0}$, and
$H_2=\frac{1}{2}d^2H/dn^2|_{n_0}$.
The density-density and density-temperature fluctuations appearing
in (\ref{p.renorm}) are difficult to compute because it is needed
to solve the fluctuating hydrodynamic equations with the full
boundary conditions imposed by the intruders. Alternatively, one
could solve the corresponding equation for the correlation
functions including the boundary conditions. A simpler estimation
of $\langle p^*\rangle$ can be obtained by employing the Fourier
transforms of the fluctuating fields $\delta \! A(\VEC r) =
V^{-1}\sum_{\VEC k} e^{-i\VEC{k}\cdot\VEC{r}} \delta \!
A_{\VEC{k}}$ and  the structure factors $S_{AB}(\VEC
k)=V^{-1}\langle \delta \! A_{\VEC{k}} \; \delta \! B_{-\VEC{k}}
\rangle$ \cite{vNE}. Expression (\ref{p.renorm}) transforms into:
\beq 
\langle p^*\rangle = p_0 + V^{-1}\sum_{\VEC k} \big[ H_1 \,
S_{nT}(\VEC k) + T_0 \, H_2 \, S_{nn}(\VEC k) \big]. \label{sum1}
\eeq

The structure factors for the uniformly driven system have been
described in detail in \cite{vNE}. The relevant contribution comes
from the region at small $k$, where they show a power law
dependence $S_{AB}(\VEC k) = S^0_{AB} \, k^{-2}$. The prefactors, $S^0_{AB}$,
depend on density, noise intensity, and restitution coefficient $\alpha$.
These asymptotic expressions, when inserted into (\ref{sum1}),
yield a sum $C \sum_{\VEC k}k^{-2}$, where the coefficient  $C \equiv
H_1 \, S^0_{nT} + T_0 \, H_2 \, S^0_{nn}$ turns out to be negative
for $n<0.73 \, d^{-2}$ and positive for larger densities.
Therefore, for $n < 0.73 \, d^{-2}$ (the case of the simulations),
fluctuations produce a decrease of the local pressure. In the gap
between the intruders (region I according to Fig.
\ref{fig.geometry}) the number of possible $\VEC k$ modes of low
$k$ is smaller than outside (region II). The effect is that the
pressure is lower outside than inside, leading to an effective
repulsive force between the intruders. Furthermore, $C$ is
proportional to the temperature, so is the Casimir force.
The sign of the force is reversed for densities  $n > 0.73 \, d^{-2}$, 
which are too close to the freezing transition or random close packing 
\cite{Tejero}
to be observed in the simulations. 
Experimentally, analogous crossover %from attractive to repulsive
is found by increasing the driving intensity~\cite{Kellay}.

%%%%%%%%%%%%%%%%% FIGURE 3 %%%%%%%%%%%%%%%%%%%%%%%%%%
%%%%%%%%%%%%%%%%% END OF FIGURE 3 %%%%%%%%%%%%%%%%%%%

Due to the long range correlations, fluctuations in regions I and
II of Fig.~\ref{fig.geometry} are correlated. However, in order to
numerically estimate the $k$-sums, we treat these regions as being
independent. This approximation will overestimate the pressure
difference and hence the Casimir force. However it provides a
rough estimation of its numerical value. In detail, we perform the
$k$-sum only over the $k$-vectors allowed by the geometrical
constraints. In a rectangular box of size $a\times b$, the
$x$-component of the $k$ vectors is $2\pi n_x/a$ and the $y$
component is $2\pi n_y/b$. It is at this point where the
difference between regions I and II appears: $a=R-D$ in region I
and $a=L-R-D$ in region II, while $b=D$ in both regions.
Moreover, the vector $\VEC k=(0,0)$ must be excluded from the sum,
and we introduce an ultraviolet cutoff, $k_c$, beyond which
hydrodynamics is not valid. We take the cutoff $k_c=2\pi/d_0$,
where $d_0=\max(d,l_0)$, and  $l_0$ is the mean free path of the
small particles.

In the limit of small $k$, with structure factors going as
$k^{-2}$, the pressure (\ref{sum1}) can be analyzed asymptotically
in the cases $a\gg b$ (for particles at large distances) and $a\ll
b$ (for particles at short distances) \beq \langle p^*\rangle =
\left\{ \begin{array}{ll}
p_0 + C\, a/b & ; \, a\gg b\gg d_0\\
p_0 + C\, b/a & ; \, b\gg a\gg d_0.\\
\end{array}\right. 
\eeq 
Note that these asymptotic expressions for the renormalized
pressure do not depend on the cutoff distance, as long as $a$ and
$b$ are much larger than it. Finally, the effective force on the
particle 2 is the difference of the forces at the left and the
right of the particle 
$ F_{2}=D\left[\langle p^*_{\rm I}\rangle
- \langle p^*_{\rm II}\rangle\right]$. 
A negative value of $C$ gives rise to a long range linear repulsive
force and a short range attractive force, at distances smaller
than $D$. The opposite is obtained when $C$ is positive.
Note that in this estimate the force depends on the system
size. This fact is related to the structure of the fluctuations,
that become larger for small wavevectors. Therefore, increasing
the system size, while keeping $R$ fixed, the fluctuations in
region II become larger, decreasing even more the pressure in this
region. However, as mentioned, at long distances this
approximation is not completely valid.

The Casimir force can be computed numerically using the full
expression of the structure factors~\cite{vNE}, not only the
$k^{-2}$ part. The computed force is shown in
Fig.~\ref{fig.force}. Note  that the linear force dependence is
preserved but the attractive part is lost, mainly because the
simulation box is finite and corrections to the $k^{-2}$ order are
observed. The computation using the $k^{-2}$ part shows a small
difference of about 10\% compared to the full calculation.
The comparison with the simulations indicates that this prediction
overestimates, as expected, the Casimir force, especially in the
$L=80d$ case, but gives the same order of magnitude and correct
sign of the force. 
Moreover the predicted force for $L=80d$ is larger 
that for $L=60d$, in agreement with the simulations. 
However, our theory does not predict the saturation of 
the force observed for $R\gtrsim 20d$. 
Possible sources for this discrepancy are: (a)
hydrodynamic correlations between regions I
and II reducing the pressure difference, 
(b) geometrical factors that arise from considering rectangular
regions instead of those bounded by circles, and (c) 
fixed intruders break the Galilean 
invariance and may modify the structure factors at very 
short wavelengths.

To summarize we have found that two intruders, immersed in a sea of smaller
granular particles driven by a white noise force, experience a
long range mutual repulsion. This repulsion has a dynamical origin
and cannot be explained by standard depletion forces. We have
proposed a mechanism based on the confinement of hydrodynamic
fluctuations when the intruders are near. The present effect is
an example of repulsion determined by fluctuation-induced forces
instead of the standard attraction; a phenomenon which has been
predicted to occur also in polymers \cite{Obukhov}. 
% The sign on
% the force has its origin in the sum over wavevectors for the
% renormalized pressure, where the (negative) density-temperature
% fluctuations give a larger contribution than the (positive)
% density-density ones. This effect is reversed when increasing the
% density beyond $n=0.73 \, d^{-2}$, leading to a attractive force.
% Experimentally, analogous crossover from attractive to repulsive
% is found by increasing the driving intensity~\cite{Kellay}.
% Besides, it is expected that different driving mechanisms or
% geometries can lead to attractive effective interactions between
% intruders, as predicted for other systems \cite{Bartolo2}.

We claim that the force we observe is the granular analog of the 
Casimir effect. We propose a novel method, valid for non-equilibrium 
fluids, to compute Casimir forces starting from the structure factors.
The two key ingredients which render this effect
manifest in the context of granular gases are: (a) the occurrence of large
low-$k$ fluctuations, originating in the coupling of nonconserving noise
with conserving hydrodynamic fields (conservation of particle number and
momentum in collisions)~\cite{vNE}, (b) the confinement of these
fluctuations in a gap (the space between the intruders), which is the common
feature of all instances of Casimir forces.
 The enhancement of the low-$k$ components plays the role of criticality in
equilibrium molecular systems, where forces are induced by the
thermal fluctuations of a correlated fluid in a confining
geometry.
Finally, these long-range forces might be responsible for
segregation effects in vibrofluidized granular mixtures of
particles having different material properties.

{\em Acknowledgments:} We acknowledge partial
support from RIKEN. 
C.C. and F.N. acknowledge JSPS, NSA and ARDA under AFOSR
Contract No.~F49620-02-1-0334, and by the NSF Grant No.~EIA-0130383.
R.B. is supported by Projects FIS04-271 (Spain) and UCM PR27/05-13923-BSCH. 
U.M.B.M. is financed by Grant Cofin-Miur 2005,
2005027808. 
R.S. is partly financed by {\em
Fondecyt} (grant 1030993), {\em Fondap} (11980002) and
Univ. Complutense (Profesores Visitantes).

\end{document}